\documentclass{revtex4}\usepackage{amssymb}
\usepackage{amsmath}
\usepackage{graphicx}
\usepackage{subfigure}

\begin{document}

\title{Two-dimensional Airy waves and three-wave solitons in quadratic media}
\author{Unchittha Prasatar$^{1}$, Thawatchai Mayteevarunyoo$^{1,4}$, and
Boris A. Malomed$^{2,3}$}
\address{$^{1}$Department of Electrical and Computer Engineering, Faculty of
Engineering, Naresuan University, Phitsanulok 65000, Thailand\\
$^{2}$Department of Physical Electronics, School of Electrical Engineering,
Faculty of Engineering, and the Center for Light-Matter University, Tel Aviv
University, Tel Aviv, Israel\\
$^{3}$Instituto de Alta Investigaci\'{o}n, Universidad de Tarapac\'{a},
Casilla 7D, Arica, Chile\\
$^4$The corresponding author}

\begin{abstract}
We address the dynamics of two-dimensional (2D) truncated Airy waves and
three-component solitons in the system of two fundamental-frequency and
second-harmonic fields, coupled by quadratic ($\chi ^{(2)}$) terms. The
system models second--harmonic-generating optical media and atomic-molecular
mixtures in Bose-Einstein condensates. In addition to stable solitons, the
system maintains truncated-Airy-waves states in either one of the
fundamental-frequency components, represented by exact solutions, which are
stable, unlike the Airy waves in the degenerate (two-component) $\chi ^{(2)}$
system. It is also possible to imprint vorticity onto the 2D Airy modes. By
means of systematic simulations, we examine interactions between truncated
Airy waves originally carried by different fundamental-frequency components,
which are bending in opposite directions, through the second-harmonic field.
The interaction leads to fusion of the input into a pair of narrow solitons.
This is opposed to what happens in the 1D system, where the interacting Aity
waves split into a large number of solitons. The interaction of truncated
Aity waves carrying identical imprinted vorticities creates an additional
pair of solitons, while opposite vorticities create a set of small-amplitude
\textquotedblleft debris" in the output. Slowly moving solitons colliding
with a heavy truncated Airy wave bounce back, faster ones are absorbed by
it, and collisions are quasi-elastic for fast solitons. Soliton-soliton
collisions lead to merger into a single mode, or elastic passage, for lower
and higher velocities, respectively.
\end{abstract}

\maketitle

\section{Introduction}

Airy waves are fundamental modes governed by linear Schr\"{o}dinger
equations, which propagate with self-acceleration along self-bending
trajectories, under the action of asymmetry in the intrinsic structure of
the modes, while the total momentum of the wave packet may be zero \cite%
{Berry}. A problem with physical realization of the Airy wave is that its
integral norm (or total power, in terms of optics) diverges. To solve this
problem, one can introduce a \textit{truncated Airy wave} (TAW), which is
also represented by an analytical solution of the linear Schr\"{o}dinger
equation, which demonstrates curvilinear propagation similar to that of the
full wave \cite{Christo}. The generation of TAWs has been predicted and
observed in diverse settings, including photonics \cite{Christo2}-\cite%
{Efremidis}, matter waves in Bose-Einstein condensates (BECs) \cite{BEC},
waves on the surface of water \cite{water}, plasmonics \cite{plasm1}-\cite%
{plasm3}, gas discharge \cite{discharge}, and electron beams \cite{el}. Much
interest has been drawn to Airy modes due to their robustness and remarkable
ability to restore themselves under the action of various perturbations, and
pass (or circumvent) obstacles. These properties open perspectives for
unique applications, such as transfer of small particles by Airy beams along
parabolic trajectories \cite{Baumgart}.

The propagation of Airy waves distorted by the self-interaction was also
studied, under the action of cubic \cite{ChrSegev-nonlin}-\cite{Conti} and
quadratic (i.e., $\chi ^{(2)}$) \cite{Ady-3wave}-\cite{we3} nonlinearities.
In the latter case, the TAW launched in the fundamental-frequency (FF)
component creates the second-harmonic (SH) one via the quadratic
upconversion. The present work addresses dynamics of two-dimensional (2D)
TAWs and solitons in a three-wave system coupled by quadratic terms, which
is also called the Type-II or nondegenerate $\chi ^{(2)}$ system \cite%
{rev1,rev2,rev3}. It includes two FF components with orthogonal
polarizations of light, and a single SH component.\

In the two-wave (degenerate) $\chi ^{(2)}$ system, which includes a single
FF wave, the dynamics of TAWs is intrinsically unstable. Indeed, 1D and 2D
TAW solutions can be easily found in the SH component, whose dynamics is
linear in the absence of the FF one. Furthermore, it is straightforward to
construct 2D TAW vortices, as combinations of products of 1D components with
a proper phase shift between them \cite{Dai,Khonina,Jiang,el,Li,Chen}.
However, any SH mode is subject to the parametric instability against small
FF perturbations. As demonstrated in works \cite{we1} and \cite{we2}, the
instability splits the linear TAW into sets of 1D or 2D solitons (which are
stable objects in the 1D and 2D $\chi ^{(2)}$ systems alike \cite{rev2,rev3}%
).

As opposed to that, in the three-wave system analytical TAW solutions
carried by either one of the FF components are stable. Also stable are
solitons, which can be found as numerical solutions of the three-component
system. These facts suggest one to consider interactions between originally
linear TAWs carried by different FF components, as well as collisions
between solitons and TAWs. Results of the systematic numerical studies of
the 1D system were reported in \cite{we3}. It was found that the TAW-TAW
interaction transforms (splits) them into a set of solitons. Furthermore,
the TAW absorbs an incident small-power soliton, and vice versa, a
high-power soliton absorbs the TAW. Between these limit cases, the collision
with a soliton splits the TAW in two solitons, or gives rise to a complex
TAW-soliton bound state.

The present work aims to produce systematic results for TAW-TAW and
TAW-soliton interactions in the 2D\ three-component system with the $\chi
^{(2)}$ nonlinearity, which is an experimentally relevant setting. The
system is formulated in Section II, which also includes the analytical
single-component TAW solution and numerical solutions for three-component
solitons. Numerical findings for the interactions are presented in Section
III. In this section we report results for TAW-TAW interactions, including
TAW pairs with identical or opposite embedded vorticities. The same section
also addresses soliton-TAW collisions. The paper is concluded by Section IV.

\section{The three-wave system}

\subsection{The basic system and TAW solution}

The standard $\chi ^{(2)}$ system of Type II for amplitudes of the optical
waves is written in the scaled form as follows \cite{rev2,rev3}:
\begin{eqnarray}
iu_{z}+bu+\frac{1}{2}\left( u_{xx}+u_{yy}\right) +v^{\ast }w &=&0,  \notag \\
iv_{z}-bv+\frac{1}{2}\left( v_{xx}+v_{yy}\right) +u^{\ast }w &=&0,
\label{type-II} \\
2iw_{z}-qw+\frac{1}{2}\left( w_{xx}+w_{yy}\right) +uv &=&0,  \notag
\end{eqnarray}%
where $u$ and $v$ are two components of the FF field, which represent
orthogonal optical polarizations, $w$ is the single SH\ field, $\ast $
stands for the complex conjugate, $z$ is the propagation distance, and $%
\left( x,y\right) $ are transverse coordinates. The Laplacian which acts on
the coordinates represents the paraxial diffraction of light. Further, real
constants $b$ and $q$ are the birefringence and $\chi ^{(2)}$ mismatch
parameter, respectively. Applying additional rescaling to Eq. (\ref{type-II}%
), one can normalize the mismatch, which makes it sufficient to consider
three values, $q=+1,0,-1$, while parameter $b$ remains irreducible.

In addition to their realization to the optical media with quadratic
nonlinearity, the same equations \ref{type-II}) apply, as coupled
Gross-Pitaevskii equations, to a mixed atomic-molecular BEC, with $z$
replaced by time and wave functions $u$ and $v$ representing two different
atomic states, while $w$ is the wave function of the molecules built as
bound states of the atomic states \cite{Drummond}. In this case, constants $%
b $ and $q$ determine, respectively, shifts of chemical potentials of the
atomic and molecular components.

The system (\ref{type-II}) conserves the total power (norm),%
\begin{equation}
P=\int \int \left( |u|^{2}+|v|^{2}+4|w|^{2}\right) dxdy,  \label{P2D}
\end{equation}%
total momentum,%
\begin{equation}
\mathbf{M}=i\int \int \left( u\nabla u^{\ast }+v\nabla v^{\ast }+2w\nabla
w^{\ast }\right) dxdy,  \label{M2D}
\end{equation}%
and angular momentum.%
\begin{equation}
\Omega =\int \int \left( u\hat{L}u^{\ast }+v\hat{L}v^{\ast }+2w^{\ast }\hat{L%
}w\right) dxdy,  \label{Om}
\end{equation}%
where the helicity operator is $\hat{L}\equiv i\left( y\partial /\partial
x-x\partial /\partial y\right) $. In the simulations presented below, the
conservation of the dynamical invariants holds with relative accuracy $\sim
10^{-6}$.

Stationary solutions to Eqs. (\ref{type-II}) are looked for in the usual
form,%
\begin{equation}
u=e^{ik_{1}z}U(x,y),v=e^{ik_{2}z}V(x,y),w=e^{i\left( k_{1}+k_{2}\right)
z}W(x,y),  \label{uvw}
\end{equation}%
where $k_{1,2}$ are real propagation constants of the FF components (in the
BEC system, with $z$ replaced by time, $-k_{1,2}$ are chemical potentials of
the atomic states), and complex functions $U\left( x,y\right) $, $V\left(
x,y\right) $ and $W\left( x,y\right) $ satisfy the following equations:%
\begin{gather}
bU+\frac{1}{2}\left( \frac{\partial ^{2}U}{\partial x^{2}}+\frac{\partial
^{2}U}{\partial y^{2}}\right) +V^{\ast }W=k_{1}U,  \notag \\
-bV+\frac{1}{2}\left( \frac{\partial ^{2}V}{\partial x^{2}}+\frac{\partial
^{2}V}{\partial y^{2}}\right) +U^{\ast }W=k_{2}V,  \label{Omega} \\
\frac{1}{2}\left( \frac{\partial ^{2}W}{\partial x^{2}}+\frac{\partial ^{2}W%
}{\partial y^{2}}\right) +UV=\left[ 2\left( k_{1}+k_{2}\right) +q\right] W.
\notag
\end{gather}%
Numerical solutions of Eq. (\ref{Omega}) which represent solitons were found
by means of the quantity-conserving squared-operator method (QCSOM) \cite%
{Yang,Yang2}. The computations were performed in the spatial domain $%
-256\leq x,y\leq +256$, discretized in each dimension by $2^{11}$ grid
points. As seen below, 2D solitons dealt with in this work have much smaller
sizes, while the TAWs are broader structures, therefore the use of the large
domain is necessary.

The analytical solution for TAW in the single FF\ component ($u$), with $%
v=w=0$, is given by the known expression \cite{Christo},%
\begin{gather}
u_{\mathrm{TAW}}\left( x,y,z\right) =u_{0}\mathrm{Ai}\left( \alpha x-\frac{%
\alpha ^{4}}{4}z^{2}+i\aleph \alpha ^{2}z\right) \mathrm{Ai}\left( \beta y-%
\frac{\beta ^{4}}{4}z^{2}+i\beth \beta ^{2}z\right)  \notag \\
\times \exp \left[ -\frac{i}{12}\left( \alpha ^{6}+\beta ^{6}\right) z^{3}+%
\frac{i}{2}\left( \alpha ^{3}x+\beta ^{3}y\right) z\right]  \notag \\
\times \exp \left[ \left( \aleph \alpha x+\beth \beta y\right) -\frac{1}{2}%
\left( \aleph \alpha ^{4}+\beth \beta ^{4}\right) z^{2}+\frac{i}{2}\left(
\aleph ^{2}\alpha ^{2}+\beth ^{2}\beta ^{2}\right) z+ibz\right] ,
\label{2D-AW}
\end{gather}%
where $\mathrm{Ai}$ is the standard Airy function, while $u_{0}$, $\alpha $,$%
\beta $ and $\aleph ,\beth $ are arbitrary positive constants. In
particular, $\aleph $ and $\beth $ determine the truncation of the Airy
wave, which makes its integral power (\ref{P2D}) convergent,

\begin{equation}
P_{\mathrm{TAW}}=\frac{u_{0}^{2}}{8\pi \sqrt{\aleph \beth }\alpha \beta }%
\exp \left( \frac{2}{3}\left( \aleph ^{3}+\beth ^{3}\right) \right) ,
\label{P2DAW}
\end{equation}%
while the linear and angular momenta, defined as per Eqs. (\ref{M2D}) and (%
\ref{Om}), are equal to zero. The TAW solution (\ref{2D-AW}) is generated by
the initial condition%
\begin{equation}
u\left( x,y,z=0\right) =u_{0}\mathrm{Ai}\left( \alpha x\right) \mathrm{Ai}%
\left( \beta y\right) \exp \left( \aleph \alpha x+\beth \beta y\right) .
\label{z=0}
\end{equation}%
Unlike a similar linear solution in the SH component of the Type-I
(two-component, alias degenerate) $\chi ^{(2)}$ system \cite{we1,we2},
solution (\ref{P2DAW}), as well as its counterpart in the $v$ component,
with $u=w=0$, are stable against small perturbations appearing in other
components.

In addition to 1D and 2D TAWs, Airy waves were recently considered in
generalized Schr\"{o}dinger equations with fractional diffraction,
characterized by the corresponding L\'{e}vy index \cite{fract1,fract2}. This
direction for the study of Airy waves may be quite promising too.

\subsection{Soliton solutions}

Equations (\ref{Omega}) give rise to exponentially localized spatial
solitons, under conditions $k_{1}>b$ and $k_{2}>-b$, in the
axially-symmetric form,
\begin{equation}
U_{\mathrm{sol}}\left( x,y\right) =U(r),V_{\mathrm{sol}}(x,y)=V(r),W_{%
\mathrm{sol}}(x,y)=W(r),  \label{r}
\end{equation}%
where $r\equiv \sqrt{x^{2}+y^{2}}$. Accordingly, Eq. (\ref{Omega}) is
transformed into a system of radial equations:
\begin{gather}
-\left( k_{1}-b\right) U+\frac{1}{2}\left( \frac{d^{2}U}{dr^{2}}+\frac{1}{r}%
\frac{dU}{dr}\right) +V^{\ast }W=0,  \notag \\
-\left( k_{2}+b\right) V+\frac{1}{2}\left( \frac{d^{2}V}{dr^{2}}+\frac{1}{r}%
\frac{dV}{dr}\right) +U^{\ast }W=0,  \label{rrr} \\
-2\left( k_{1}+k_{2}\right) W-qW+\frac{1}{2}\left( \frac{d^{2}W}{dr^{2}}+%
\frac{1}{r}\frac{dW}{dr}\right) +UV=0.  \notag
\end{gather}%
A typical profile of three components of a soliton is displayed in Fig. \ref%
{fig3}. It was obtained starting from input
\begin{equation}
U\left( x,y\right) =2.5\mathrm{sech}\left( 0.8r\right) ,V(x,y)=2.2\mathrm{%
sech}(0.8r),W(x,y)=1.9\mathrm{sech}(0.8r).  \label{input}
\end{equation}%
Stability of this solution has been verified by direct simulations of its
perturbed evolution, using the split-step algorithm. While we did not aim to
develop the stability analysis for the full family, all the solitons dealt
with in this work are completely stable objects.

\begin{figure}[tbp]
\centering{\includegraphics[width=3.0in]{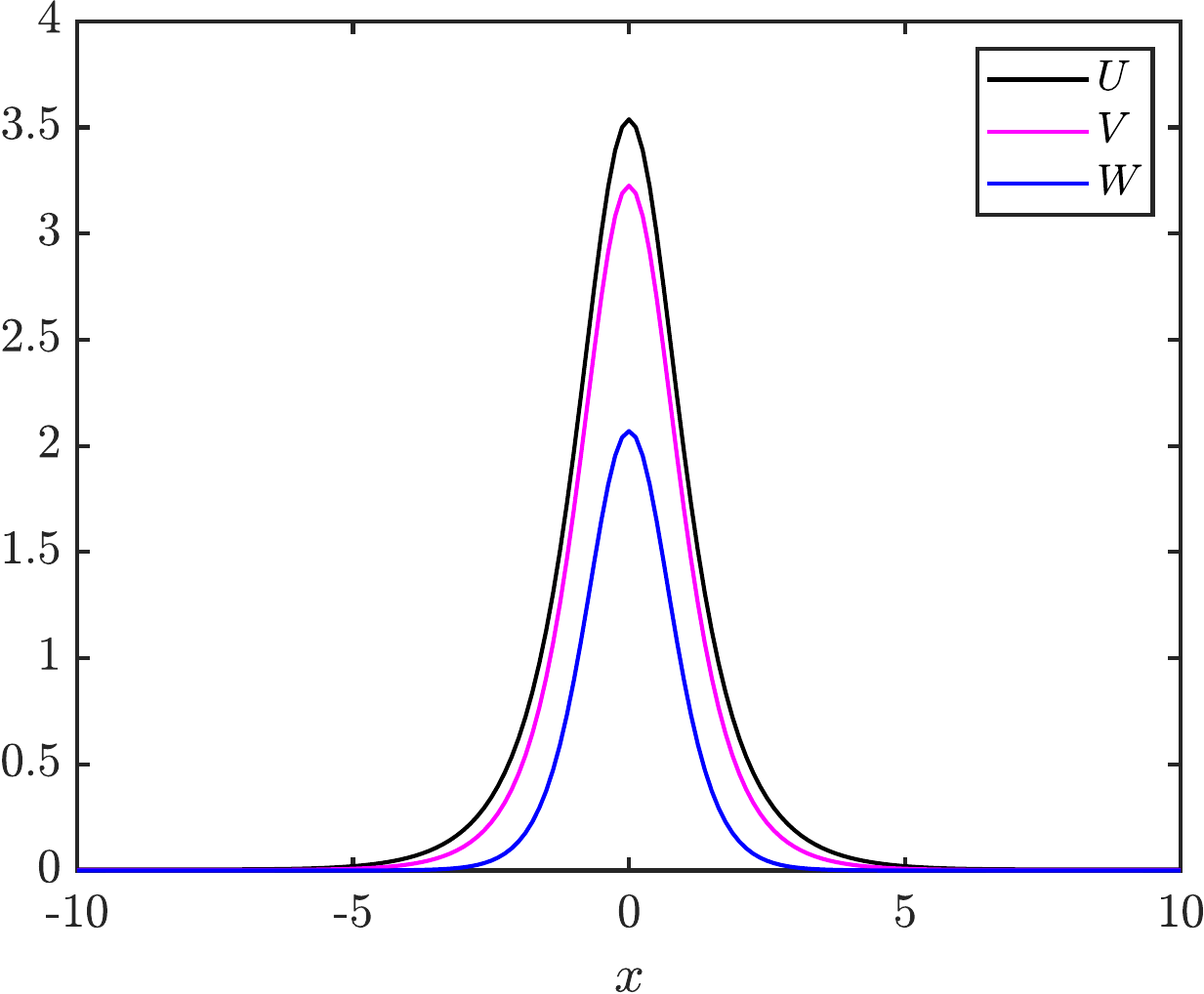}}
\caption{Profiles $U(x)$, $V(x)$, and $W(x)$ of a stable axisymmetric
soliton in cross section $y=0$, obtained as a numerical solution of Eq. (%
\protect\ref{rrr}) with $k_{1}=0.6444,k_{2}=1.0763$, $b=0,q=+1$. The total
power of this soliton, defined as per Eq. (\protect\ref{P2D}), is $P=100$.}
\label{fig3}
\end{figure}

\begin{figure}[tbp]
\centering\includegraphics[width=5in]{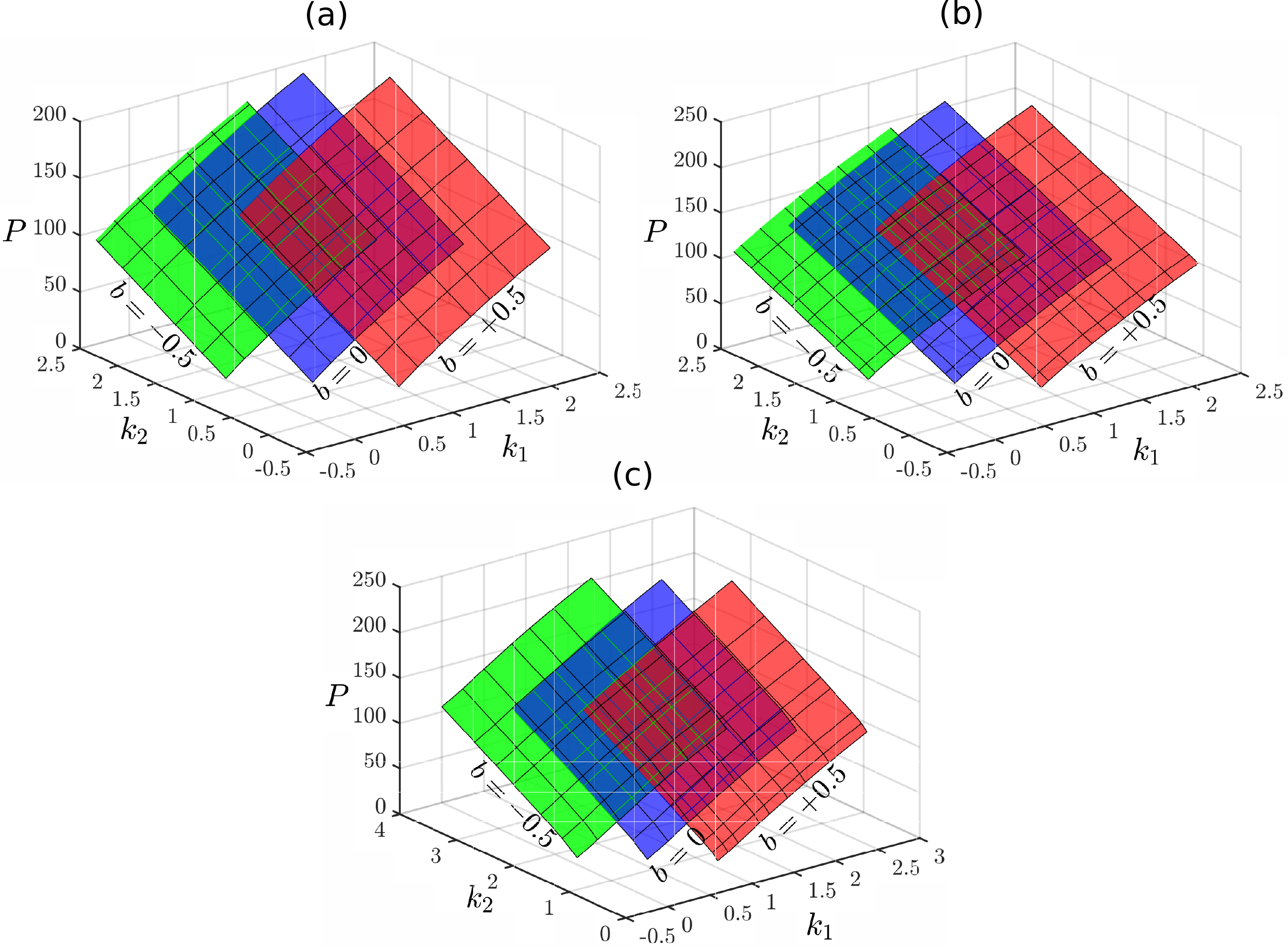}
\caption{The total power (\protect\ref{P2D}) of solitons, $P(k_{1},k_{2})$,
produced by the numerical solution of Eqs. (\protect\ref{rrr}), for
different values of birefringence $b$, as indicated in the panels, and
different values of the mismatch: (a) $q=0$, (b) $q=+1$, and (c) $q=-1$.}
\label{fig1}
\end{figure}

Families of stable soliton solutions are characterized by the dependence of
the total power, $P$, on the propagation constants, $k_{1,2}$, which is
displayed in Fig. \ref{fig1}, for fixed values of the mismatch: (a) $q=0$;
(b) $q=+1$; (c) $q=-1$. In each panel, colors represent different values of
the birefringence: blue for $b=0$, red for $b=+0.5$, and green for $b=-0.5$.

Two-dimensional solitons of the three-wave system with embedded vorticity
are available too, but they all were found to be unstable against
spontaneous splitting into fundamental (zero-vorticity) modes. The latter
result, which may be construed as the modulational instability of a vortex
ring against transverse perturbations, is not reproduced here, as it was
demonstrated for $\chi ^{(2)}$ systems previously \cite{Torner,Mihalache}.

Because the underlying equations (\ref{type-II}) maintain the Galilean
invariance, tilted solitons can be obtained from the straight ones by the
application of the Galilean boost with components $\left( c,d\right) $:%
\begin{eqnarray}
u(x,y,z) &=&u_{\mathrm{sol}}(x-x_{0}-cz,y-y_{0}-dy)\exp \left[ i\left(
cx+dy\right) -\frac{i}{2}\left( c^{2}+d^{2}\right) z\right] ,~  \notag \\
v(x,y,z) &=&v_{\mathrm{sol}}(x-x_{0}-cz,y-y_{0}-dy)\exp \left[ i\left(
cx+dy\right) -\frac{i}{2}\left( c^{2}+d^{2}\right) z\right] ,~  \label{sol}
\\
w(x,y,z) &=&w_{\mathrm{sol}}(x-x_{0}-cz,y-y_{0}-dy)\exp \left[ 2i\left(
cx+dy\right) -i\left( c^{2}+d^{2}\right) z\right] .  \notag
\end{eqnarray}%
In these expressions, $\left( x_{0},y_{0}\right) $ is the position of the
soliton's center at $z=0$.

\begin{figure}[tbp]
\centering\includegraphics[width=4in]{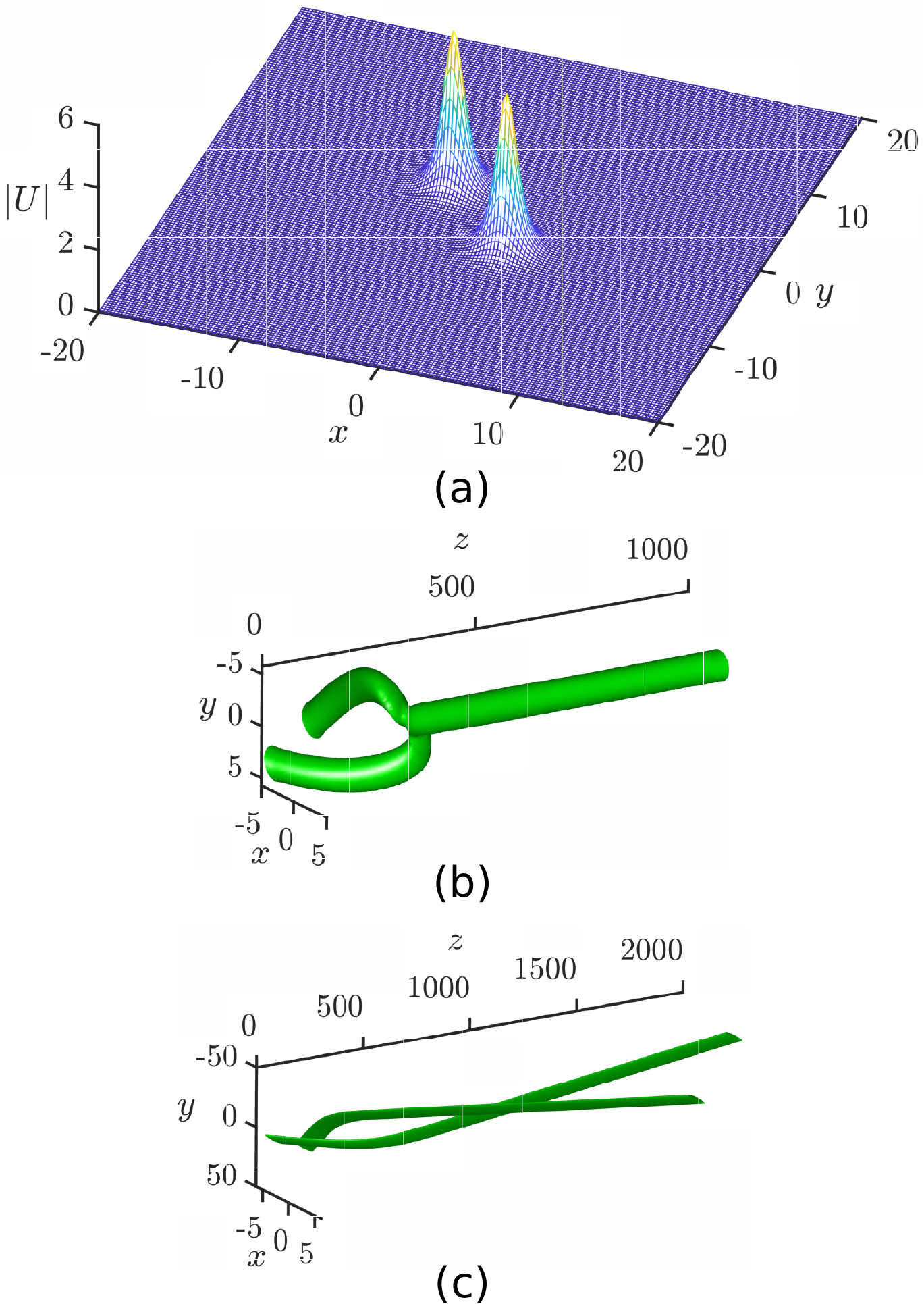}
\caption{(a) The two-soliton stationary bound state, numerically found in
the reference frame rotating with angular velocity $\protect\omega =0.01$.
This solution corresponds to propagation constants $k_{1}=0.7304$, $%
k_{2}=1.2773$ of the two FF components, while the total power is $P=227.14$.
The mismatch and birefringence parameters are $q=+1$ and $b=0$. Panels (b)
and (c) demonstrate instability of the two-soliton bound state by means of
isosurfaces of $|u\left( x,y;z\right) |$, produced by direct simulations in
the case when the distance between the two solitons is made, respectively,
slightly smaller or larger than the equilibrium value given by Eq. (\protect
\ref{ll}).}
\label{fig-two-sol}
\end{figure}

\begin{figure}[tbp]
\centering\includegraphics[width=4in]{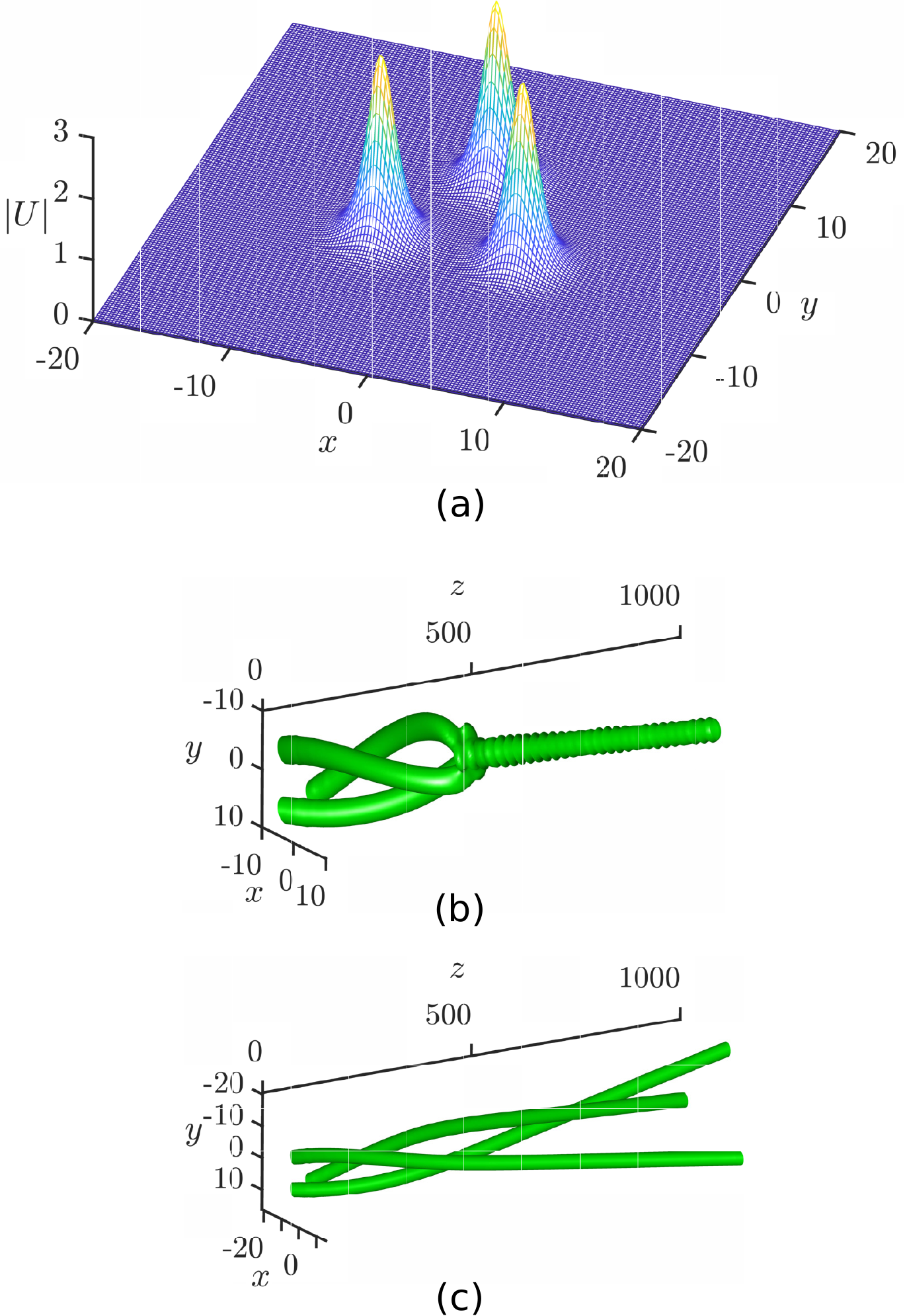}
\caption{(a) The three-soliton stationary bound state, numerically found in
the reference frame rotating with angular velocity $\protect\omega =0.01$.
This solution corresponds to propagation constants $k_{1}=0.40083$, $%
k_{2}=0.7139$ of the two FF components, while the total power is $P=211.12$.
The mismatch and birefringence parameters are $q=+1$ and $b=0$. Panels (b)
and (c) demonstrate instability of the three-soliton bound state by means of
isosurfaces of $|u\left( x,y;z\right) |$, produced by direct simulations in
the case when the distance between the two solitons is made, respectively,
slightly smaller or larger than the equilibrium value given by Eq. (\protect
\ref{ll}).}
\label{fig-three-sol}
\end{figure}

It is relevant to attempt constructing a \textit{rotating bound state} of
two identical solitons, or three ones in the form of a rotating equilateral
triangle, assuming that attraction between in-phase solitons \cite%
{attraction} is in balance with the centrifugal force. This can be done
looking for stationary two- and three-soliton solutions in the reference
frame rotating with angular velocity $\omega $. The result is that such
states with $\omega $ small enough can be produced, but they all are
unstable. These results are illustrated by Figs. \ref{fig-two-sol} and \ref%
{fig-three-sol}, respectively. The two- and three-soliton rotating bound
states with parameters corresponding to Figs. \ref{fig-two-sol} and \ref%
{fig-three-sol} are found with the equilibrium distance between solitons
being, respectively,
\begin{equation}
\left( l_{\mathrm{eq}}\right) _{\mathrm{2-sol}}\approx 5.926;~\left( l_{%
\mathrm{eq}}\right) _{\mathrm{3-sol}}\approx 11.936,  \label{ll}
\end{equation}%
as shown in panels (a) of the figures. If the perturbation makes the
distance slightly smaller than $l_{\mathrm{eq}}$, the solitons merge into a
single one, while the perturbation which makes the distance slightly larger
than $l_{\mathrm{eq}}$ splits the bound state into sets of gradually
separating solitons, as seen in panels (b) and (c), respectively, of Figs. %
\ref{fig-two-sol} and \ref{fig-three-sol}.

\section{Collisional dynamics in the three-wave system}

\subsection{Soliton-soliton collisions}

First, it is relevant to examine interactions between stable solitons in the
three-component system. Here we focus on head-on collisions between
identical solitons, generated by Eq. (\ref{sol}) under conditions%
\begin{equation}
\left( x_{0}=y_{0}\right) _{1}=-\left( x_{0}=y_{0}\right) _{2}>0,\left(
d=c\right) _{1}=-\left( d=c\right) _{2}>0,  \label{==}
\end{equation}%
where subscripts $1$ and $2$ pertain to the two solitons.

\begin{figure}[tbp]
\centering\includegraphics[width=4in]{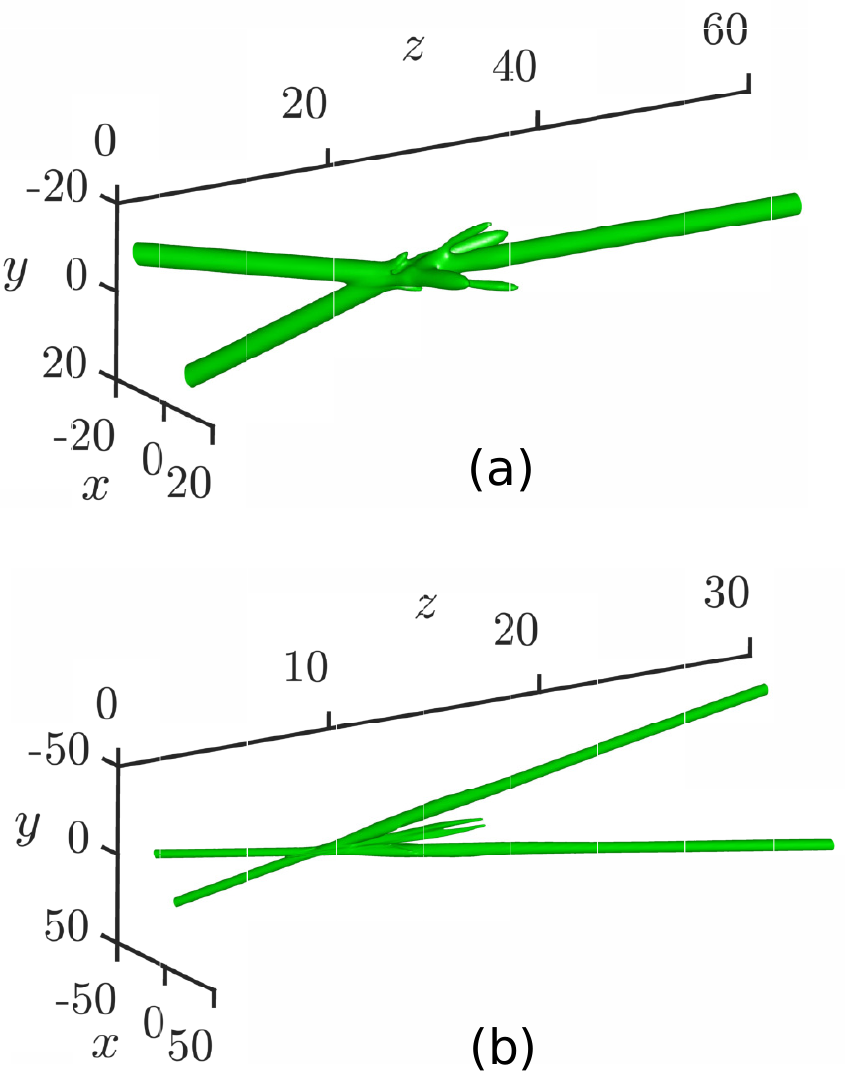}
\caption{Isosurface plots of $\left\vert u\left( x,y;z\right) \right\vert $
illustrating head-on collisions of identical solitons solitons, with the
total power $P=100$, as produced by simulations of Eq. (\protect\ref{type-II}%
) with $q=1,b=0.5$. The solitons are set in motion (in fact, tilted in the
spatial domain) as per Eq. (\protect\ref{==}) with $c_{1,2}=\pm 0.5$ (a),
and $c_{1,2}=\pm 1.5$ (b).}
\label{fig5}
\end{figure}

Systematic simulations of Eqs. (\ref{type-II}) have demonstrated that slowly
colliding solitons merge into a single quiescent one, while fast solitons
collide quasi-elastically, separating with the original values of the tilt
(with an \textquotedblleft attempt" to create an additional quiescent
soliton, which, however, decays). Typical examples of the merger and
quasi-elastic collisions are displayed, severally, in Figs. \ref{fig5}(a)
and (b).

\begin{figure}[tbp]
\centering\includegraphics[width=4in]{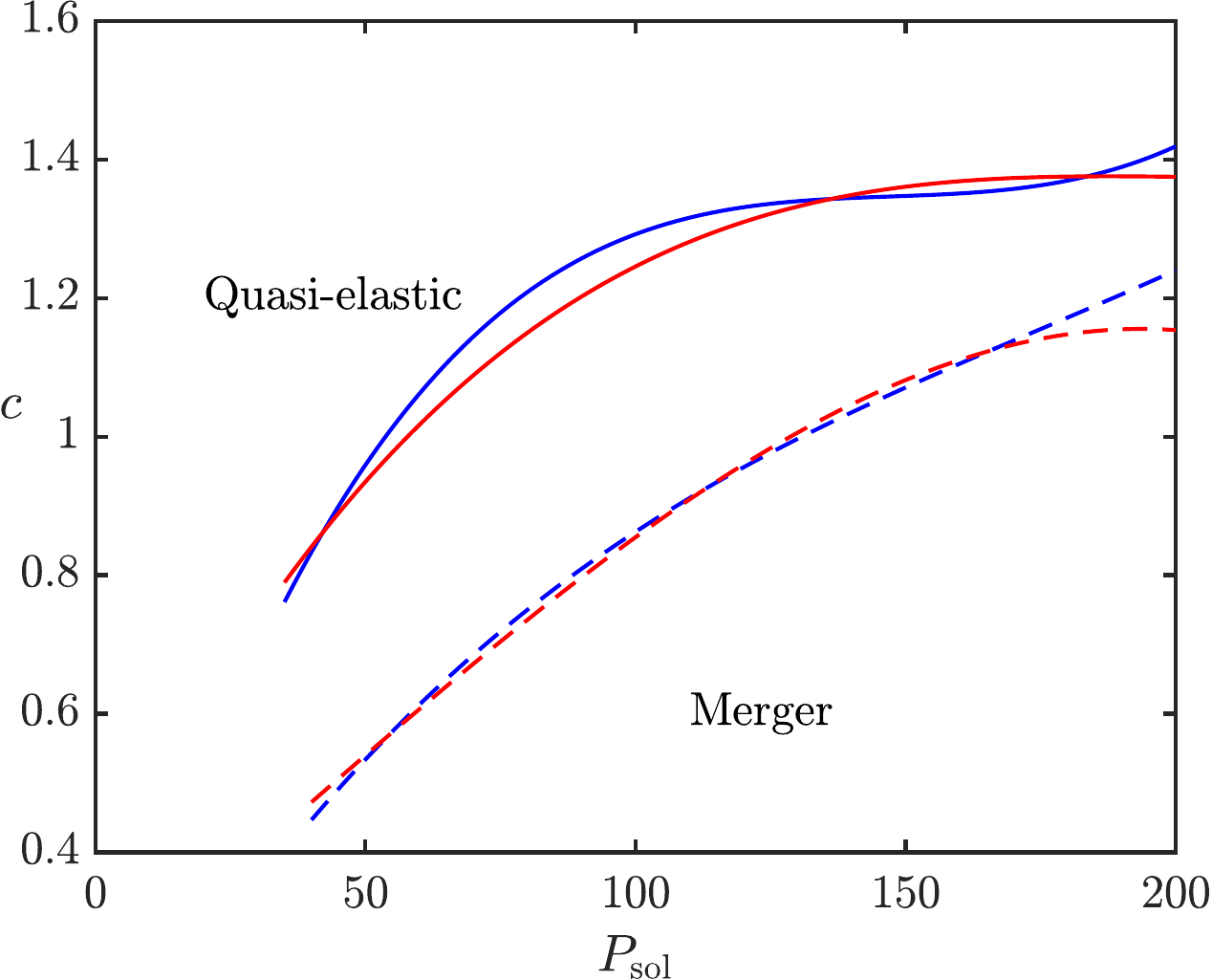}
\caption{The border in the plane of the total power, $P_{\mathrm{sol}}$, and
kick strength $c$ (see Eq. (\protect\ref{==}), between quasi-elastic
collisions and collision-induced merger of the three-component solitons. The
results are shown for two values of the mismatch, $q=+1$ and $-1$
(continuous and dashed lines, respectively), at two values of the
birefringence coefficint, $b=0$ and and $0.5$ (blue and red lines,
respectively).}
\label{fig4}
\end{figure}

The results are summarized in Fig. \ref{fig4}, which shows boundaries
between the merger and quasi-elasticity regimes in the plane of $\left(
P,c\right) $, assuming that the tilt parameters in Eq. (\ref{==}) are taken
as $c_{1,2}=d_{1,2}=\pm c$. The chart in Fig. \ref{fig4}\ demonstrates
essential dependence of the character of the collisions on the system's
mismatch, $q$, and weak dependence on the birefringence, $b$.

\subsection{Airy -- Airy collisions}

\begin{figure}[tbp]
\centering\includegraphics[width=4in]{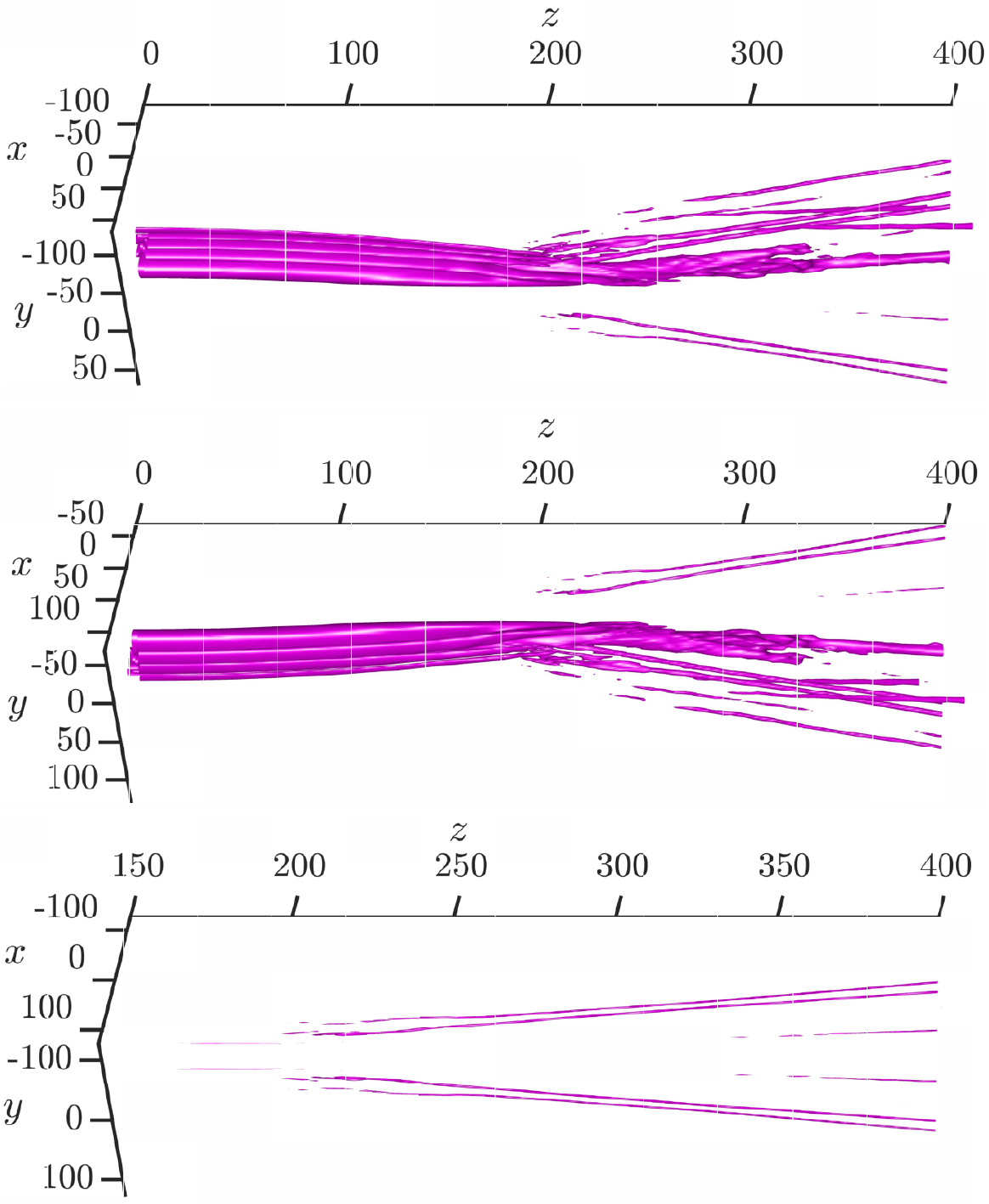}
\caption{Isosurface plots of $\left\vert u\left( x,y;z\right) \right\vert $,
$\left\vert v\left( x,y;z\right) \right\vert $, and $\left\vert w\left(
x,y;z\right) \right\vert $ illustrating the collision of the TAWs, initiated
by input (\protect\ref{AA}) with $u_{0}=10$, $\protect\alpha =\protect\beta %
=-0.12$, $\aleph =\beth =0.125$, and separation distance $2x_{0}=2y_{0}=20$.
The top, middle, and bottom panels represent the $u$-, $v$- and $w$%
-components, respectively. The collision leads to fusion of the
three-component wave field into two separated narrow solitons, to which some
\textquotedblleft debris" is added. This outcome is additionally illustrated
by Fig. \protect\ref{fig9}.}
\label{fig8}
\end{figure}

To simulate collisions between two TAWs carried by the $u$- and $v$%
-components of the FF field, initial conditions were taken as two mirror
copies of the wave form (\ref{2D-AW}), with equal amplitudes $u_{0}$ and
opposite values of $\alpha $ and $\beta $. The waves are initially separated
by distances $2x_{0}$ and $2y_{0}$ along the $x$ and $y$ axes (we set $%
x_{0}=y_{0}$):%
\begin{eqnarray}
u\left( x,y,z=0\right) &=&u_{0}\mathrm{Ai}\left( \alpha (x+x_{0})\right)
\mathrm{Ai}\left( \beta (y+y_{0})\right) \exp \left( \aleph \alpha
(x+x_{0})+\beth \beta (y+y_{0})\right) ,  \notag \\
v\left( x,y,z=0\right) &=&u_{0}\mathrm{Ai}\left( -\alpha (x-x_{0})\right)
\mathrm{Ai}\left( -\beta (y-y_{0})\right) \exp \left( -\aleph \alpha
(x-x_{0})-\beth \beta (y-y_{0})\right) ,  \label{AA} \\
w\left( x,y,z=0\right) &=&0,  \notag
\end{eqnarray}%
cf. Eq. (\ref{z=0}). While each wave form in isolation is an exact solution,
the overlapping between them initiates nonlinear interaction through term $%
uv $ in the third equation of system (\ref{type-II}). The opposite signs of $%
\alpha $ and $\beta $ imply that the two initial TAWs bend in opposite
directions, as can be seen in Fig. \ref{fig8} (the bend is small in the
figure, as displaying it in a more pronounced form would require to run the
simulations in a huge domain, while the size of the bend does not strongly
affect the result of the collision).

\begin{figure}[tbp]
\centering\includegraphics[width=4in]{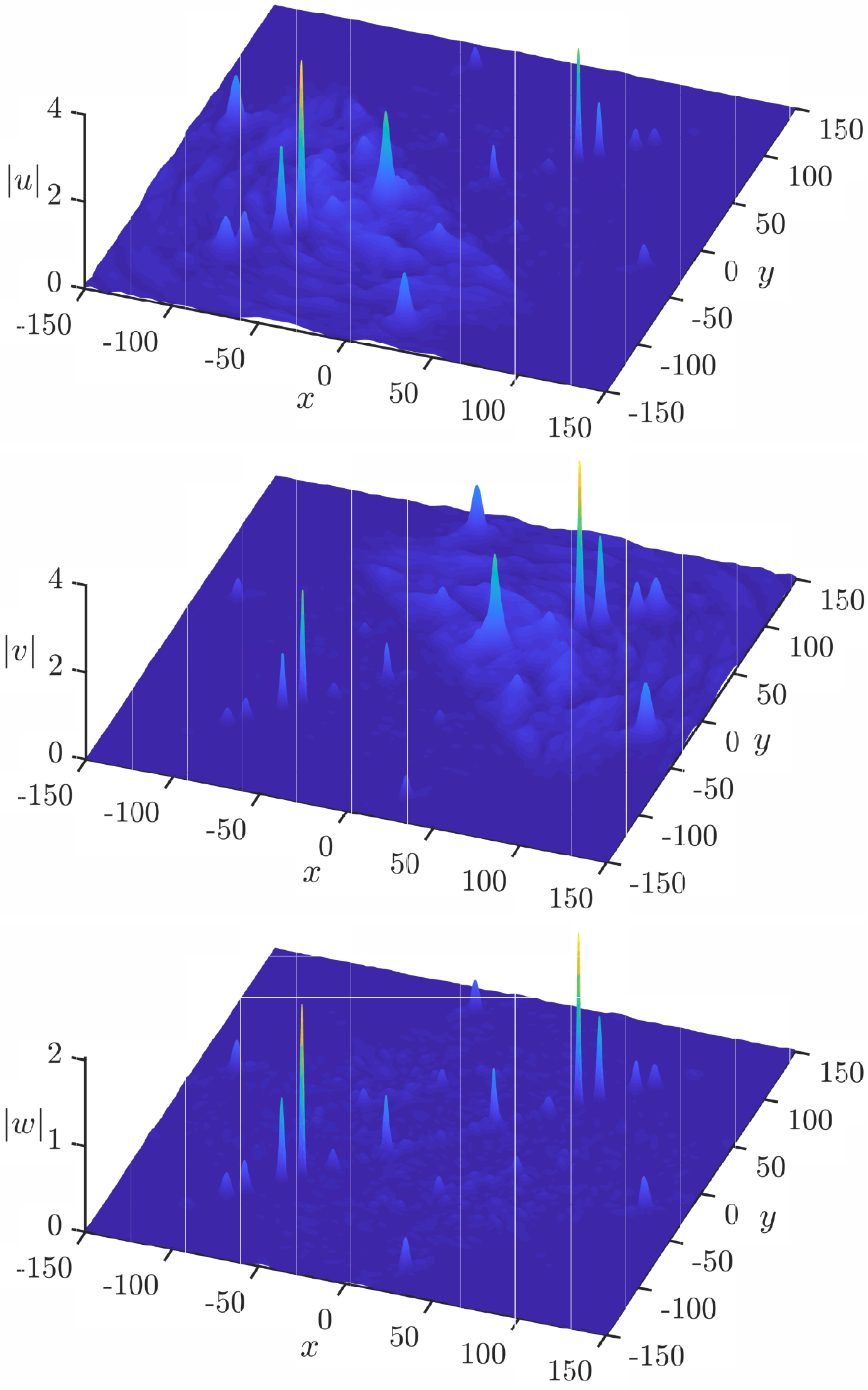}
\caption{Final amplitude profiles of the fields at $z=400$, corresponding to
Fig. \protect\ref{fig8}. The top, middle and bottom panels represent the $u$%
-, $v$- and $w$-components, respectively. }
\label{fig9}
\end{figure}

A typical overall picture of the interaction in the space of $\left(
x,y;z\right) $ and the corresponding final configuration are displayed,
respectively, in Figs. \ref{fig8} and \ref{fig9}. The outcome of the
interaction is relatively simple: fusion of the colliding broad TAWs into a
pair of well-separated narrow solitons with equal powers, accompanied by
generation of conspicuous \textquotedblleft debris". It is relevant to
mention that, in the 1D $\chi ^{(2)}$ system, considered in work \cite{we3},
collision of TAWs also lead to creation of a set of solitons,{\huge \ }but
the number of emerging solitons was much larger -- typically, between $7$\
and $12$, with different powers of individual solitons. In fact, in the 1D
setting the collision splits the colliding multi-peak Airy waves, each major
peak giving rise to a soliton. On the contrary, in the 2D case the
simulations actually reveal not splitting but \emph{fusion} of the field
into just two solitons. This observation is naturally explained by the
stronger self-focusing effect in 2D. Indeed, in the cascading limit, which
corresponds to large mismatch \cite{rev3}, the three-wave $\chi ^{(2)}$\
system may be reduced to a system of coupled nonlinear Schr\"{o}dinger
equations with the cubic self-focusing nonlinearity, which, by itself, leads
to the 2D collapse \cite{Berge,Sulem,Fibich}. Of course, the collapse is not
eventually reached, as the cascading approximation eventually breaks down,
but the trend to strong self-focusing is obvious in the 2D setting.

\subsection{Collisions between Airy vortices}

To consider the dynamics including TAW with imprinted vorticity (helicity),
the simplest possibility is to add helical factors, $\exp \left(
im_{u,v}\theta \right) $, with integer winding numbers $m_{u.v}$, to the
input given by Eq. (\ref{AA}) ($\theta $ is the angular coordinate in the $%
\left( x,y\right) $ plane). Thus, the accordingly modified input is%
\begin{eqnarray}
u\left( x,y,z=0\right) &=&u_{0}\mathrm{Ai}\left( \alpha (x+x_{0})\right)
\mathrm{Ai}\left( \beta (y+y_{0})\right) \exp \left( \aleph \alpha
(x+x_{0})+\beth \beta (y+y_{0})\right) \exp \left( im_{u}\theta \right) ,
\notag \\
v\left( x,y,z=0\right) &=&u_{0}\mathrm{Ai}\left( -\alpha (x-x_{0})\right)
\mathrm{Ai}\left( -\beta (y-y_{0})\right) \exp \left( -\aleph \alpha
(x-x_{0})-\beth \beta (y-y_{0})\right) \exp \left( im_{v}\theta \right) ,
\label{vort} \\
w\left( x,y,z=0\right) &=&0.  \notag
\end{eqnarray}%
Strictly speaking, this ansatz does not represent an exact TAW-vortex
solution, but it corresponds to the experimentally relevant situation, in
which vorticity is imparted to an available beam, such as TAW, by passing it
through an appropriately designed helical phase plate \cite{plate,plate2} or
computer-generated hologram \cite{holo,KH}.

\begin{figure}[tbp]
\centering\includegraphics[width=4in]{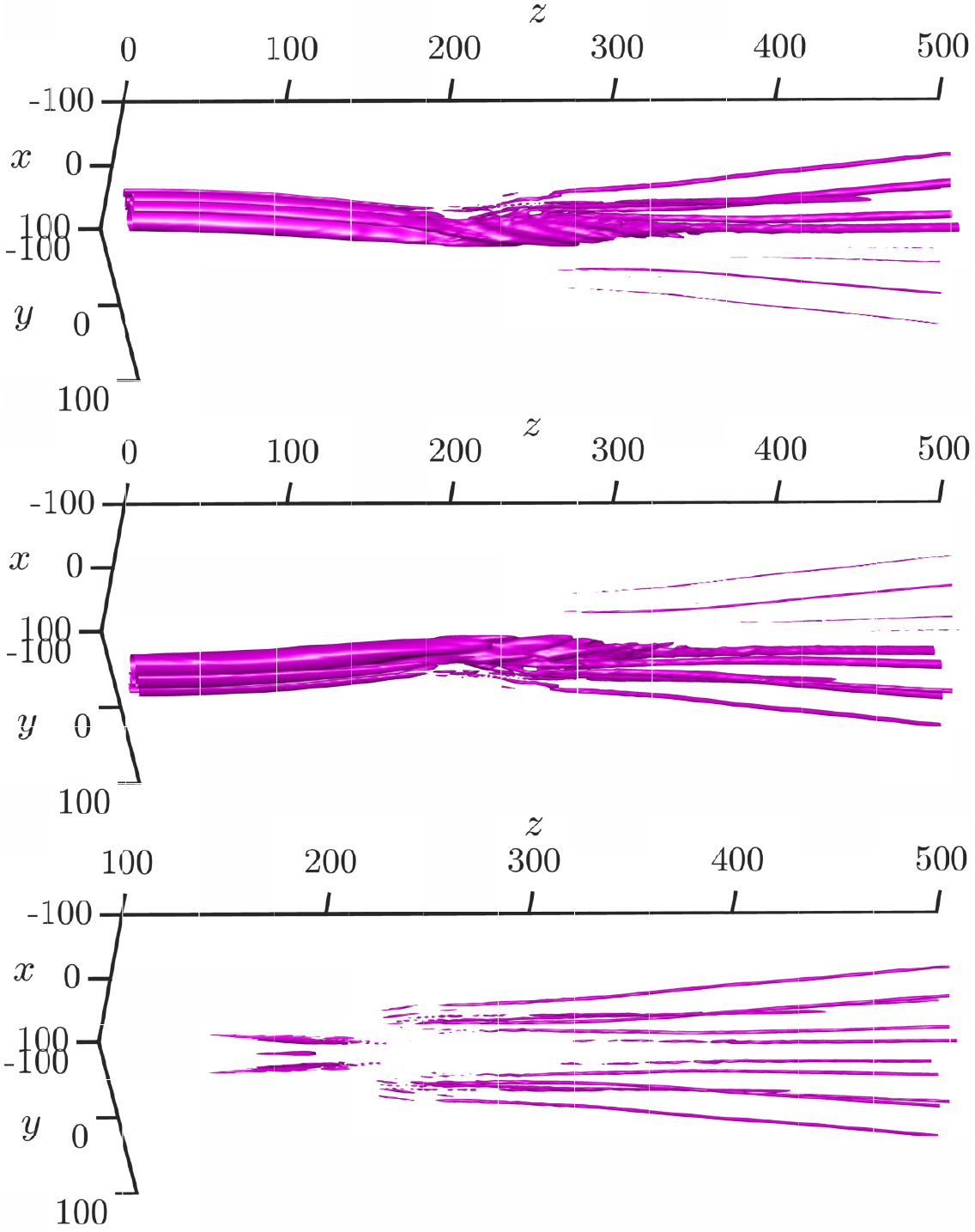}
\caption{The same as in Fig. \protect\ref{fig8}, but for the collision of
TAWs with superimposed vorticity, initiated by input (\protect\ref{vort})
with equal winding numbers, $m_{u}=m_{v}=1$. The parameters are $u_{0}=10$, $%
\protect\alpha =\protect\beta =-0.12$, $\aleph =\beth =0.125$, and $%
2x_{0}=2y_{0}=20$.}
\label{fig11}
\end{figure}

\begin{figure}[tbp]
\centering\includegraphics[width=4in]{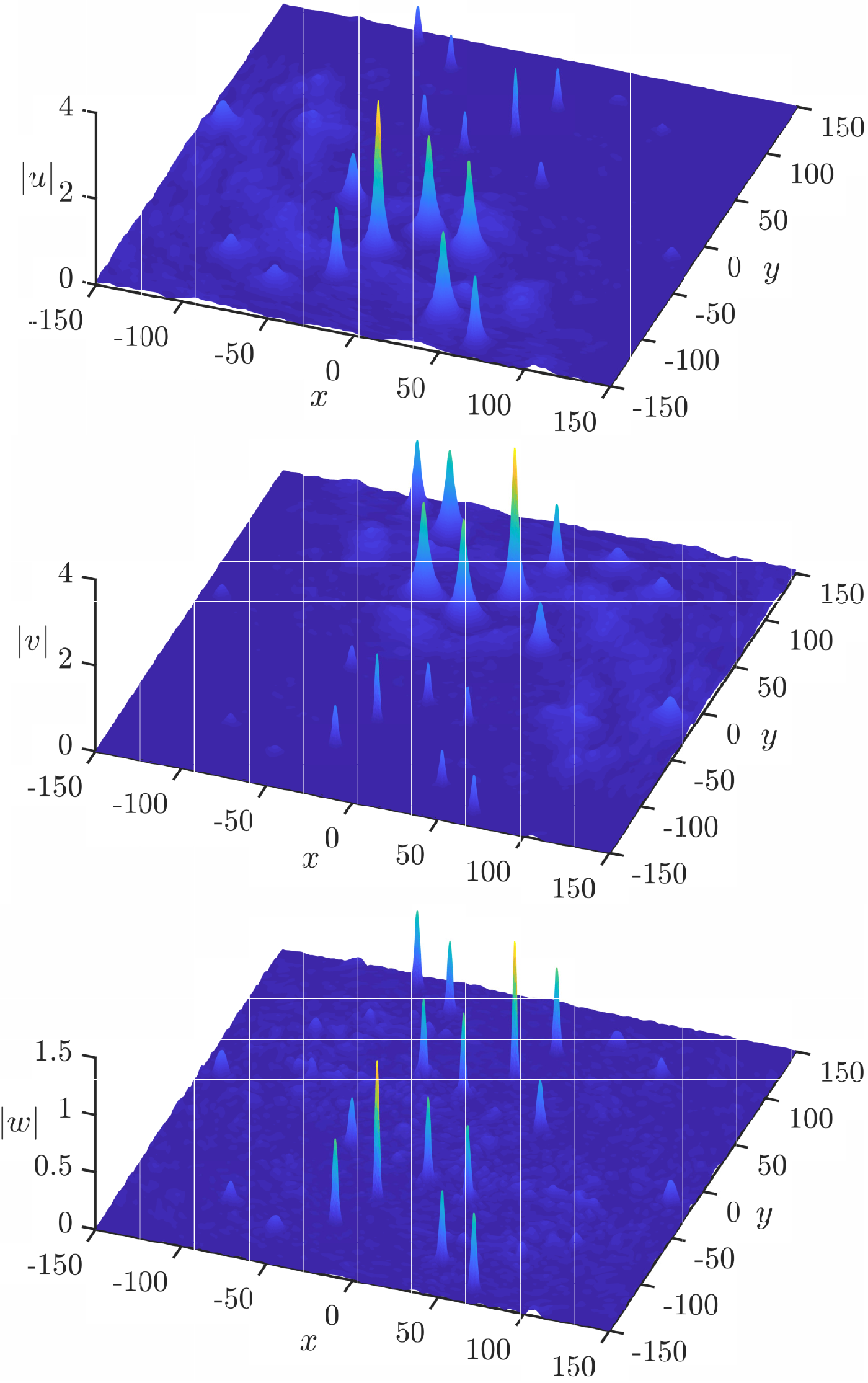}
\caption{The final amplitude profiles at $z=500$ corresponding to Fig.
\protect\ref{fig11}. The top, middle and bottom panels display the $u$-, $v$%
- and $w$-components, respectively. }
\label{fig12}
\end{figure}

The nonlinear interaction between colliding helical TAWs is affected by the
above-mentioned splitting instability of vortex solitons \cite%
{Torner,Mihalache}, which acts, as a transient effect, on the helical Airy
modes as well. As a result, the collision converts the interacting vortical
TAWs into a pair of well-separated narrow solitons, similar to what is
demonstrated above in Figs. \ref{fig8} and \ref{fig9}, to which additional
solitons some \textquotedblleft debris" are added. A typical example of the
collision of the pair of mirror-symmetric TAWs with identical winding
numbers, $m_{u}=m_{v}=1$, is displayed in Figs. \ref{fig11} and \ref{fig12}.

\begin{figure}[tbp]
\centering\includegraphics[width=4in]{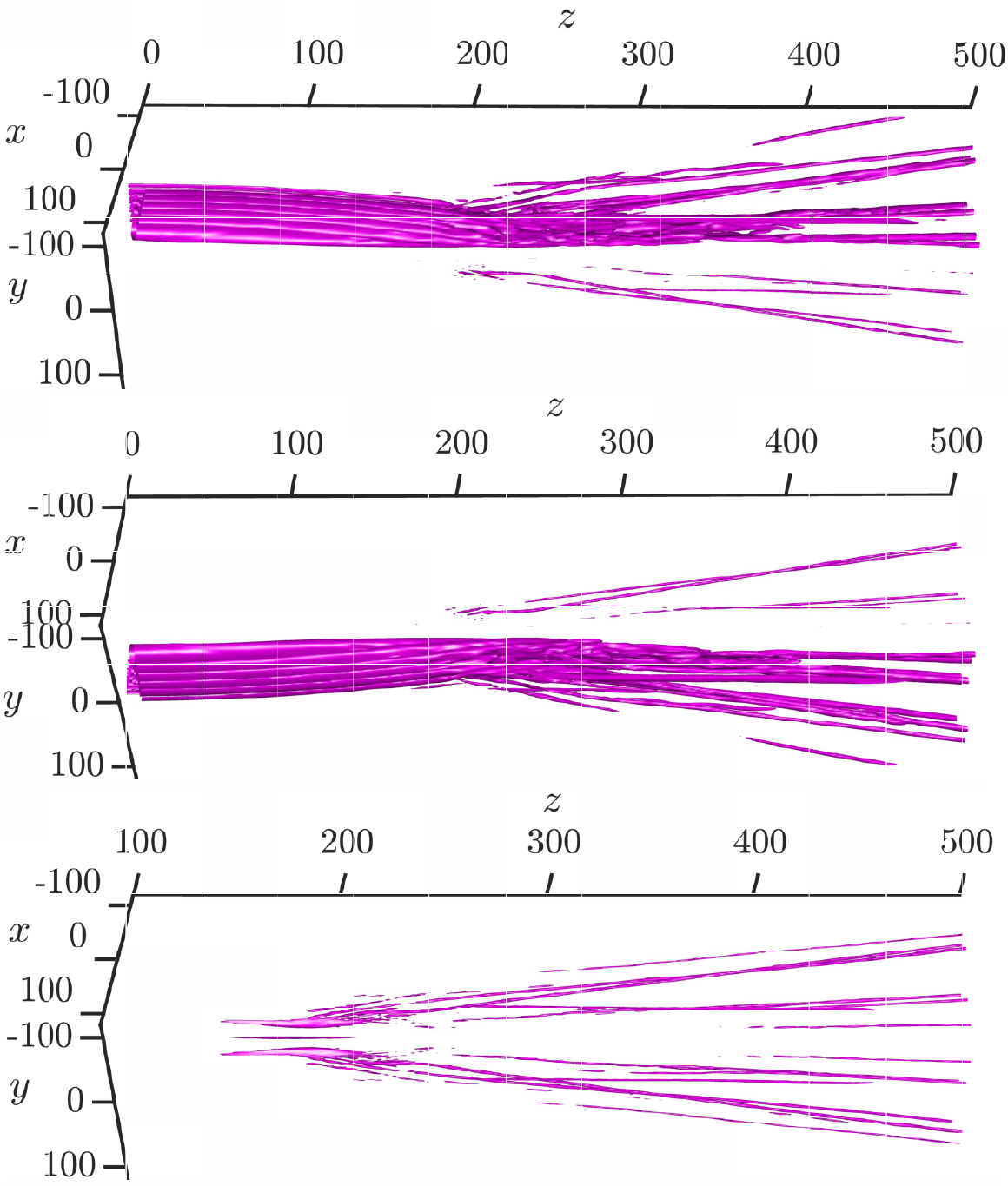}
\caption{The same as in Fig. \protect\ref{fig11}, but for the collision of
Airy vortices with opposite vorticities, \textit{viz}., $\left(
m_{u},m_{v}\right) =(+1,-1)$ in Eq. (\protect\ref{vort}). The parameters are
$u_{0}=10$, $\protect\alpha =\protect\beta =0.12$, $\aleph =\beth =0.125$,
and $x_{0}=y_{0}=50$.}
\label{fig13}
\end{figure}

\begin{figure}[tbp]
\centering\includegraphics[width=4in]{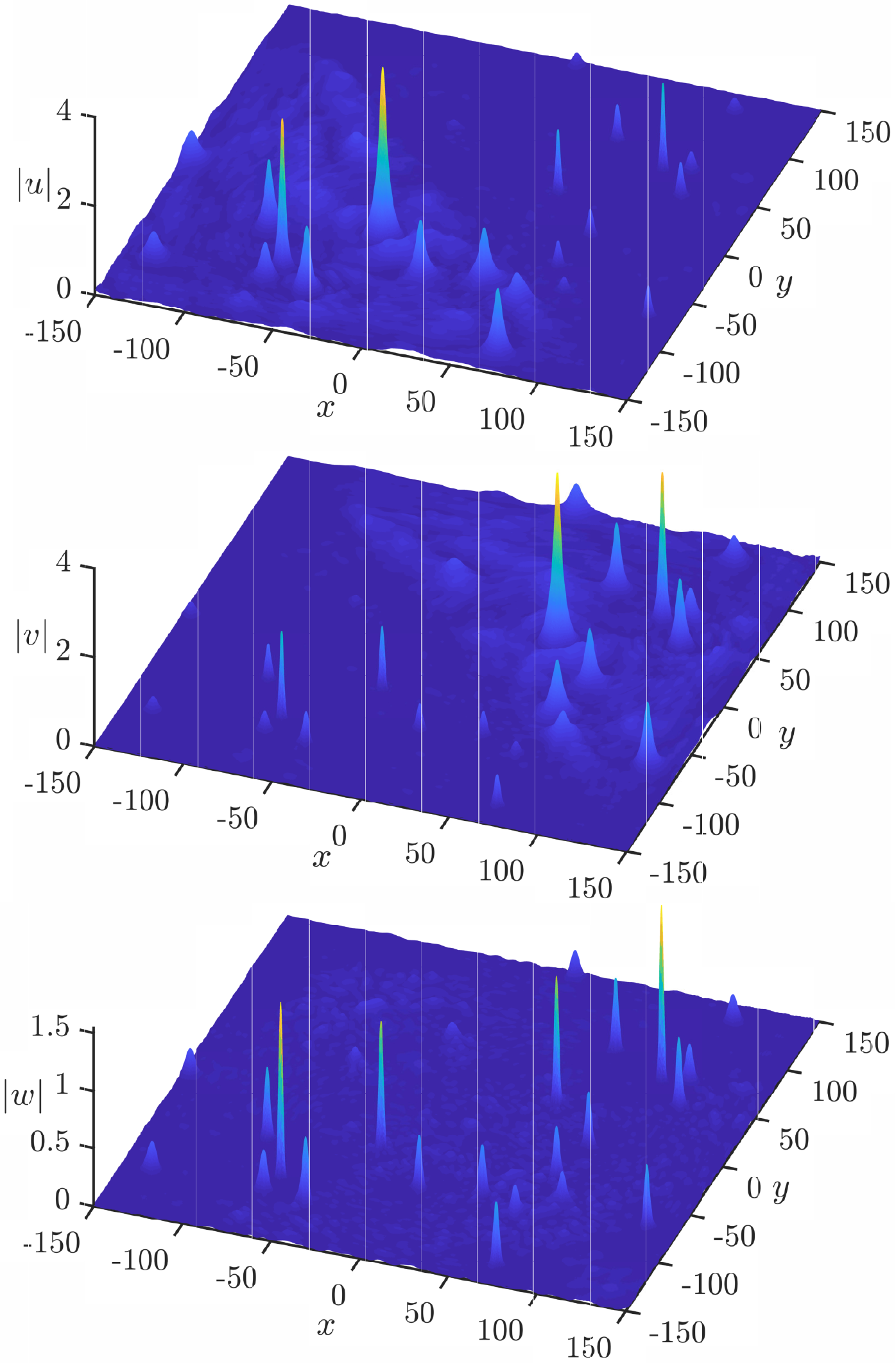}
\caption{Final amplitude profiles at $z=400$ corresponding to Fig. \protect
\ref{fig13}. The top, middle and bottom panels display the $u$-, $v$- and $w$%
-components, respectively. }
\label{fig14}
\end{figure}

Further, a typical example of the collision between TAWs carrying opposite
vorticities, i.e., $m_{u}=-m_{v}=1$, is presented by Figs. \ref{fig13} and %
\ref{fig14}. In this case, the dominant dynamical factor is not the
splitting instability of vortex rings, but the Kelvin-Helmholtz instability
of counterpropagating azimuthal flows. Manifestations of this instability
for optical vortices are known in nonlinear photonic media \cite{KH}. It
tends to produce a larger number of separated fragments. Accordingly, Figs. %
\ref{fig13} and \ref{fig14} demonstrate the creation of a pair of main
narrow solitons, along with a cluster of small-amplitude fragments.

\section{Airy -- soliton collisions}

Proceeding to collisions of solitons with TAWs, it is natural to consider
the case when a relatively light soliton impinges upon a heavy Airy wave.
Here, we present results for fixed parameters of the TAW,%
\begin{equation}
\alpha =\beta =0.12,\aleph =\beth =0.125,  \label{parameters}
\end{equation}%
whose center is placed at $x_{0}=y_{0}=-50$, see the first line in Eq. (\ref%
{AA}). The amplitude of the incident soliton was taken to be equal to the
TAW's amplitude:
\begin{equation}
\left( \left\vert u_{\mathrm{sol}}\right\vert \right) _{\max }=u_{0}.
\label{uu}
\end{equation}
The soliton's center was originally placed at $\left( x_{0}\right) _{\mathrm{%
sol}}=\left( y_{0}\right) _{\mathrm{sol}}=-70$, and tilt $c$ was applied to
it as per Eq. (\ref{sol}). Data of the simulations were collected by varying
$|c|$ from $0$ to $\approx 3.5$, and varying the soliton's power between $P_{%
\mathrm{sol}}=50$ and $200$.

\begin{figure}[tbp]
\centering\includegraphics[width=4in]{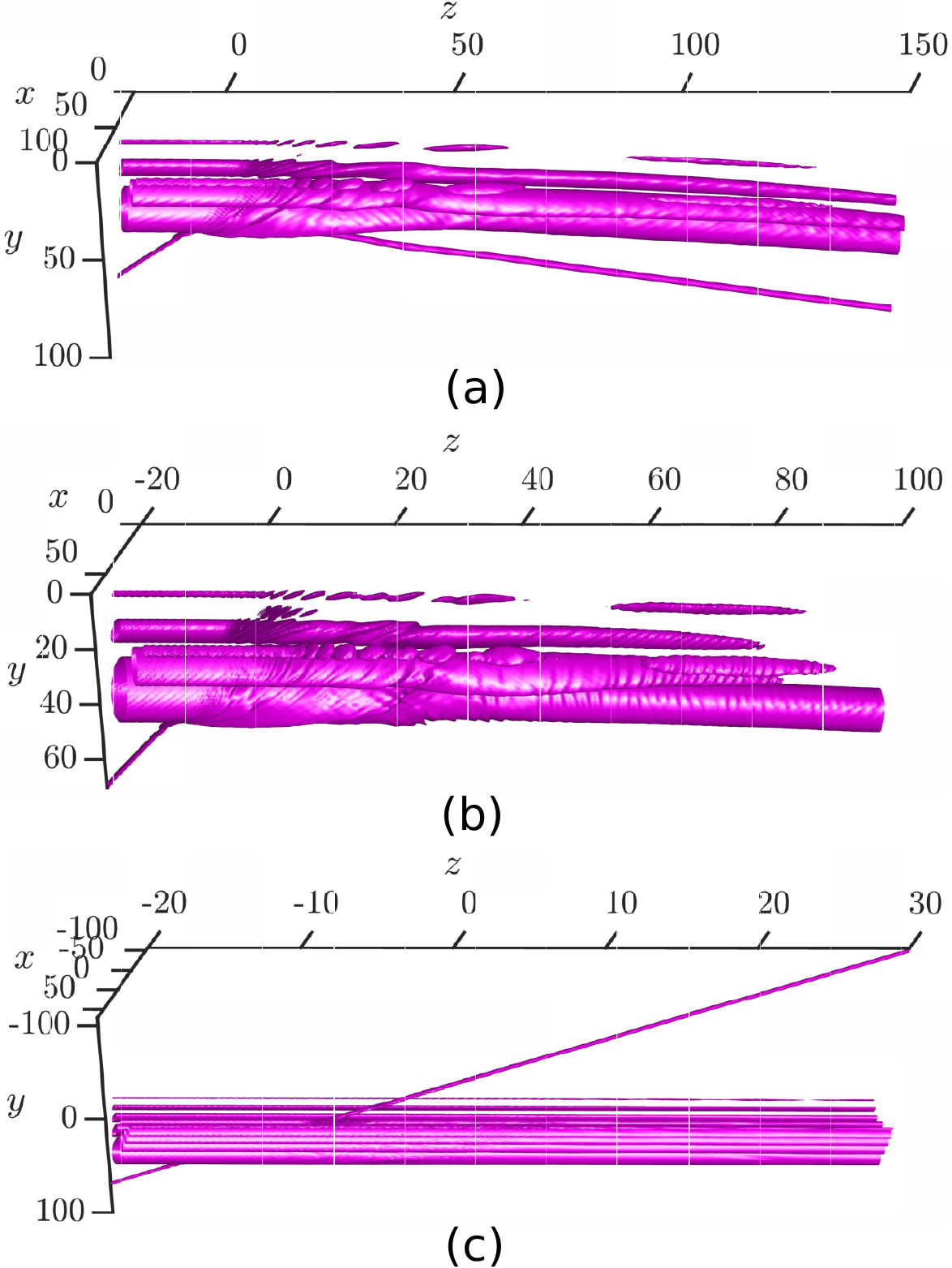}
\caption{Isosurface plots of $\left\vert u\left( x,y;z\right) \right\vert $
illustrating collisions between the soliton with power $P_{\mathrm{sol}}=100$%
, corresponding to $k_{1}=1.1765$ and $k_{2}=0.5829$, and TAW with
parameters fixed as per Eqs. (\protect\ref{parameters}) and (\protect\ref{uu}%
), other constants being $q=1$ and $b=0.5$. The tilt is $c=-1.2$ in (a), $%
c=-2.0$ in (b), and $c=-3.5$ in (c). In panel (c) self-bending of the Airy
wave is not visible, as the displayed interval of the propagation distance
is too small for that.}
\label{fig6}
\end{figure}

As shown in Fig. \ref{fig6}(a), relatively slow solitons, with $|c|\leq 1.4$%
, bounce back from the TAW. Note that rebound of the incident soliton was
not observed in the 1D three-component system \cite{we3}, where a light
soliton was always absorbed by the TAW. In the present system, the
absorption takes place in an interval of $1.4\leq |c|\leq 2.6$, see Fig. \ref%
{fig6}(b). The soliton bouncing from the TAW or merging into it produces a
visible perturbation, which, however, does not destroy the TAW. Finally, at $%
|c|\geq 2.6$, the fast soliton quickly passes the TAW, as seen in Fig. \ref%
{fig6}(c).

\begin{figure}[tbp]
\centering\includegraphics[width=4in]{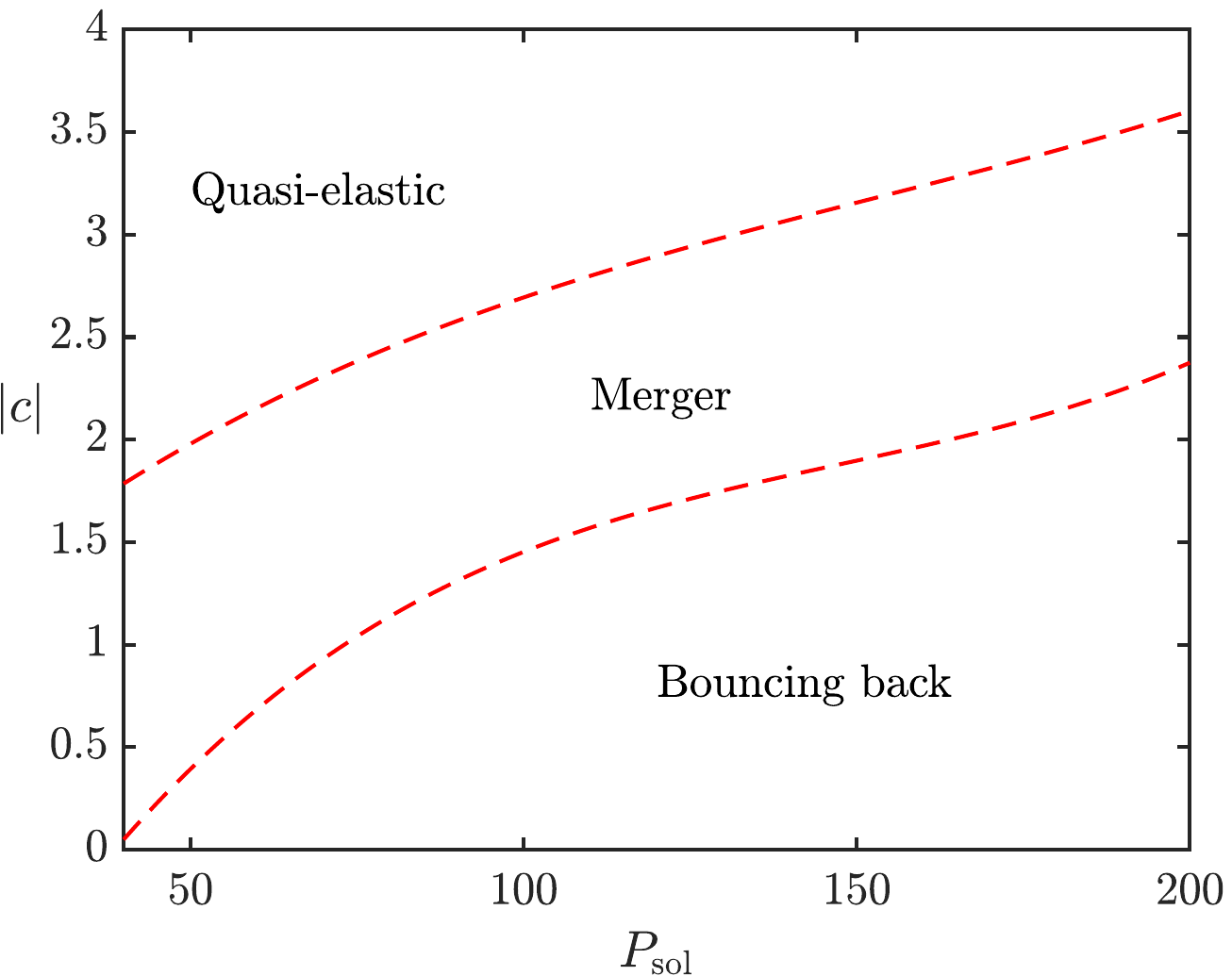}
\caption{Borders in the plane of $\left( P_{\mathrm{sol}},|c|\right) $
separating different outcomes of the soliton-TAW collisions. TAW parameters
are fixed as per Eqs. (\protect\ref{parameters}) and (\protect\ref{uu}),
other constants being $q=1$ and $b=0.5$.}
\label{fig7}
\end{figure}

The results for the soliton-TAW collisions are summarized in Fig. \ref{fig7}%
, in the plane of $\left( P_{\mathrm{sol}},|c|\right) $. The general
dependence of the critical value of the tilt constant, which is the boundary
between the rebound and merger, on the power can be easily explained by the
scaling relation between the tilt and total power, following from Eq. (\ref%
{type-II}), if the birefringence and mismatch terms are neglected (i.e., the
scaling is relevant for sufficient large values of the power): $c\sim \sqrt{P%
}$.

Simulations of the collision of the incident soliton with the vortical TAW,
generated as per input (\ref{vort}), produce results which are not
essentially different from those displayed in Figs. \ref{fig6} and \ref{fig7}%
.

\section{Conclusion}

The present work aims to address specific dynamical effects demonstrated by
TAWs (truncated Airy waves) and solitons in the 2D three-wave system of two
FF (fundamental-frequency) and SH (second-harmonic) fields coupled by the $%
\chi ^{(2)}$ terms in the optical medium. A unique peculiarity of the system
is that TAW states, carried by a single FF component, which are represented
by exact analytical solutions, are stable against small perturbations in the
form of seeds of the other components. The same system give rise to stable
three-component solitons, which can be found in the numerical form. These
facts suggest possibilities to consider interactions between TAWs carried by
the different FF components (they interact nonlinearly through the SH
field), as well as collisions between solitons and TAWs. Previously, these
possibilities were explored in the 1D system, while the present work
addresses the 2D setting. In particular, the present system makes it
possible to consider interactions involving TAWs onto which vorticity is
imprinted.

First, the interaction of two TAWs, originally created as mirror images of
each other in the different FF\ components, leads to the fusion of the
fields into a pair of well-separated narrow solitons. The character of the
interaction is opposite to that in the 1D three-wave system, where, instead
of the \textit{fusion}, the interaction leads to \textit{fission} of the
colliding TAWs into a large number of secondary solitons. The interaction of
2D\ TAWs with imprinted vorticity is affected by the transient splitting
instability of vortex rings, in the case when both TAWs carry the identical
vorticity. In this case, the interaction generates an additional pair of
solitons. In the case of opposite vorticities of the interacting TAWs, there
appears a cluster of fragments, produced by the transient Kelvin-Helmholtz
instability of the counter-rotating vortices. Further, the systematic
numerical study of collisions of relatively light solitons with a heavy TAW
demonstrates that a slowly moving soliton bounces back, a faster one is
absorbed by the TAW, while fast collision is quasi-elastic. As concerns
soliton-soliton collisions, it is shown here that they lead to merger of
slow solitons into a single one, or quasi-elastic passage of fast solitons.

A challenging possibility for the development of the analysis is to extend
it to the three-dimensional system governing the spatiotemporal propagation
of waves in $\chi ^{(2)}$ media, where solitons may be stable too \cite%
{Rub,HH}. In particular, one may expect that the interaction of 3D Airy
waves may be more violent because the cubic nonlinear Schr\"{o}dinger
equations, to which the $\chi ^{(2)}$ system reduces in the cascading limit
gives rise to the supercritical collapse \cite{Berge,Sulem,Fibich}.

\section*{Acknowledgment}

We thank Prof. Jianke Yang for sharing codes for the application of the
QCSOM numerical algorithm. This work was supported, in part, by the RGJ PhD
program (PHD/0078/2561), the Thailand Research Fund (grant No. BRG 6080017),
and the Israel Science Foundation (grant No. 1286/17).

\end{document}